\newenvironment{namelist}[1]{%
\begin{list}{}
    {
      
      \settowidth{\labelwidth}{#1}
      \setlength{\leftmargin}{1.1\labelwidth}
    }
  }{%
\end{list}}

\documentstyle[
times, twoside, epsf,
csit
]{article}
\author
    {\large
     Larissa Ismailova, Sergey Kosikov, Konstantin Zinchenko,\\
       Alexey Mikhaylov,
     Lioubouv Bourmistrova, Anastassiya Berezovskaya
     \vspace{1.52mm} \\
{\normalsize Vorotnikovsky per, 7, bld. 4} \\
{\normalsize Dept.\ for\ Advanced\ Computer\ Studies\ and\
             Information\ Technologies}\\
{\normalsize Institute for Contemporary Education ``JurInfoR-MSU''} \\
{\normalsize Moscow, 103006, Russia} \\
{\normalsize {\tt larisa@jurinfor.ru;\ lyui@jmsuice.msk.ru}}
    \date{}
    }
\title{\bf Building Views with Description Logics in ADE:\\
           Application Development Environment %
}

\institution{}

\begin{document}
\bibliographystyle{alpha}

\markboth{Building Views with Description Logics in ADE:
           Application Development Environment}
{Workshop on Computer Science and Information Technologies CSIT'2000,
Ufa, Russia, 2000}

\maketitle


\begin{abstract}
\noindent Any of views is formally defined within description
logics that were established as a family of logics for modeling
complex hereditary structures and have a suitable expressive
power. This paper considers the Application Development
Environment (ADE) over generalized {\em variable} concepts that
are used to build database applications involving the supporting
{\em views}. The front-end user interacts the database via
separate ADE access mechanism intermediated by view support. The
variety of applications may be generated that communicate with
different and distinct desktop databases in a data warehouse. The
advanced techniques allows to involve remote or stored procedures
retrieval and call.
\end{abstract}


\section*{Introduction} \label{intro}

Recent research activity generated not only the valuable advance in
the area of supporting views~\cite{BeLeRo:97}
but stimulated the experimental efforts in
developing the {\em views} supporting mechanisms
over the generalized object-oriented structures, e.g.
BACK, CLASSIC, CRACK, FLEX, K-REP, KL-ONE, KRIS, LOOM, YAK.

Here is briefly discussed the Application Development Environment
(ADE) that is used to build database applications involving the
generalized views. They are encircled within the {\em description
logics} that are a family of logics being developed for modeling a
diversity of hierarchical structures. The main units in a
description logic (DL) are the unary predicates, called {\em
concepts} or, more generalized, {\em variable
concepts}~\cite{Wo:96}. Other kind of units is represented by the
binary predicates called {\em roles}, or {\em cases}. Variable
concepts represent the sets of generalized objects (called {\em
individuals}) and cases or roles represent their states.

The generic concepts and roles are used as {\em initial primitive}
units while the additional concepts and roles are defined
using {\em constructors} giving rise to {\em derived} units.
Thus, the concepts indicate the associated classes of
{\em individual} instances in the domain and the constructors
over the concepts indicate the appropriate necessary and sufficient
conditions on individuals of the classes.

The restricted types of DL's with (ordinary) concepts
have the semantics based on the
first order logic with equality while the DL in use here
manipulates the variable concepts, thus, involving the
higher order logics and structures. This is more general
assumption than usually, e.g., the known result for
DL's with ordinary ({\em not} variable) concepts is that
it can be translated into a special kind of
first order logic~\cite{Bor:96}.

To solve the possible inconveniences, ADE computes separately
the database access and the user interaction with
this computational environment.
The variety of applications may be generated that communicate
with different and distinct desktop databases. The advanced techniques
allows to involve remote or stored procedures retrieval and call.

According to an object-oriented traditions \cite{OMG1:91},
ADE include some basic features
of inheritance, encapsulation, and polymorphism.
They are used to derive an
{\em actual} object from {\em possible}, or {\em potential} objects
to cover  the needed information resources.

The {\em potential object} (PO) is composed with the {\em menu} (M),
{\em data access} (DA),
and {\em modular counterparts} (MC).
The {\em Ancestor Potential Objects} (APO) contain
the menus, events, event evolver, attributes and functions (that are
encapsulated). The {\em Descendant Potential Objects} (DPO) are
inherited from APO.

The aim of the current contribution is to give a brief profile of
the ADE project in general and yet without any detailed
mathematical or implementation consideration. Nevertheless, some
mathematical background corresponds to the references
\cite{OMG1:91}, \cite{HeSa:95}, \cite{And:96}. Other less
traditional for the database area ideas are due to \cite{Wo:96} to
conform the {\em object computation} strategies. The main ADE
building blocks have the relative uniformity to resolve the
modular linkages. ADE enables the host computational environment
to extend the properties of the distinct MC.


To support this, we develop a well-modularized architecture
for DL that is implemented using the ``normalize-compare''
approach (see Section~\ref{integrating-data}).
This architecture expects a set of procedures to be filled in
for each new concept constructor extending the
original language.
In addition, the methodological heuristics could be proposed
to specifying what these procedures need to do.

The outline of the rest of the paper is as follows.
Section~\ref{basic-notions} outlines the basic notions
in use. Section~\ref{pr-with-concepts}
provides an introduction to
DLs, their syntax and semantic description,
and the services provided by the
reasoning with concepts, especially the ``subsumption'' relationship.
Section~\ref{event-driven-objects} provides the basics
from event-driven technics in use.
Section~\ref{integrating-data}
gives a brief outline of data integrating facilities.
It introduces the architecture of the proposed approach
to DL support,
provides an overview
of the methodology for extending it.
Section~\ref{supporting-technologies} describes the
general features of supporting technologies.
It terminates by discussing successes and limitations
of the proposed ADE approach to
extension, an its relationship to one particular
other approach that is directly relevant.

We conclude by
summarizing the contributions and limitations
of the approach.

\section{Basic notions and
         architectures for database interoperations}\label{basic-notions}

In this paper the term `view' is used in a rather general sense.
For instance, the SQL statement `{\tt CREATE VIEW}' results in
a single and {\em virtual relation} with a content being specified
using a query over pre-specified relations.
Here the term `view'
means a database schema, where the contents of the schema
elements (relations, classes, etc.) are specified using queries
against one or more already­specified schemas.
A view may
be {\em virtual} or {\em actual} (materialized),
or supported using a combination of the two.

There are several {\em known} architectures for
database interoperation: mediation, federation,
mediation with updates, workflow.

In our approach the computational environment identifies three
main layers to supporting data integration. The first of them
holds {\em wrappers} supporting common query interface. The second
one holds {\em mediators} and provides semantic data integration.
At last, the third layer holds {\em coordinators} providing
support for relevant information sources.

\section{Preliminaries with concepts}\label{pr-with-concepts}

Description logics are knowledge representation languages
tailored for expressing knowledge about concepts and concept
hierarchies. They are usually given a Tarski style declarative
semantics, which allows them to be seen as sub-languages of
predicate logic.
They are considered an important formalism unifying and giving
a logical basis to the well known traditions
of {\em frame-based} systems,
semantic networks and KL-ONE-like languages,
object-oriented representations, semantic data models,
and type systems.
The {\em basic building blocks} are {\em concepts}, {\em roles}
and {\em individuals}.

{\em Concepts} describe the common properties of a collection of
individuals and can be considered as unary predicates
which are interpreted as sets of individuals.

{\em Roles} are interpreted as binary relations between individuals.
Each description logic defines also a number of language
constructs (such as intersection, union, role quantification, etc.)
that can be used to define new concepts and roles.
The main reasoning tasks are classification and satisfiability,
subsumption and instance checking.
Subsumption represents the is-a relation.
Classification is the computation of a concept hierarchy based on
subsumption. A whole family of knowledge representation
systems have been built using these languages
and for most of them complexity results for the main reasoning
tasks are known. Description logic systems have been used for
building a variety of applications including conceptual modeling,
information integration, query mechanisms, view maintenance,
software management systems, planning systems,
configuration systems, and natural language understanding.

\subsection{Example}

Consider an example with the {\em primitive concepts} {\tt
person}, {\tt technical} and {\tt
In\-for\-ma\-tion\-Tech\-nologies}, and {\tt paper} is a {\em
primitive role}. The new derived concepts can be defined by the
various constructors. This is a way to imposing some restrictions
on the number of fillers of a certain role have to be associated.
E.g., {\tt ($\le$ 6 paper)} is a derived concept determined by
applying a {\em number} restriction on the role {\tt paper}.
Namely, it determines the set of objects that have at most 6
fillers for the case (role) {\tt paper}. The same way, {\tt
($\forall$ paper.technical)} is a description that derives the
concept, -- an {\em intension}, -- by imposing the {\em universal
quantifier} constructor $\forall$, and the associated class -- an
{\em extension}, -- returns the class of objects with the property
that all the fillers of the case {\tt paper} are within the class
{\tt technical}. {\tt ($\exists$
paper.In\-for\-ma\-tion\-Tech\-nologies)} contains the {\em
existential quantifier} and determines the set of objects, -- {\em
individuals}, -- among which there exists at least one filler of
the case {\tt paper} belonging to the class {\tt
In\-for\-ma\-tion\-Tech\-nologies}.
The following concept description describes the complex concept
{\tt C1}. This concept includes a set of persons such that
all of their papers are technical, have at most 6 published papers,
and have a paper that is on Information Technologies area:
\begin{center}
{\tt C1 := person $\sqcap$ ($\forall$ paper.technical) $\sqcap$ \\
            ($\le$ 6 paper) $\sqcap$ \\
        ($\exists$ paper.InformationTechnologies)     }
\end{center}
If we want to fix the set of {\em primary facts}, then
the assertions like
\begin{center}
{\tt Concept(Instance)}, \\
{\tt Role(Owner,Member)}
\end{center}
are to be used, where {\tt Instance}, {\tt Owner} and {\tt Member}
are the {\em individuals}.

For instance, {\tt paper(Rick,`Logics in Hu\-ma\-ni\-ties')}
means that {\tt `Logics in Hu\-ma\-ni\-ties'} is Rick's paper.
{\tt ($\forall$ paper.technical)(Rick)} means that all the
Rick's papers are technical, i.e. belong to the class
{\tt technical}.

\subsection{Syntax and semantics}

Now we determine the outline for syntax and semantics
of description logic used in this paper.

\subsubsection{Syntax}

In the definition below $A$ denotes a primitive concept,
$P_i$'s are used for indicating the roles, $C$, $C_1$ and $C_2$
denote concept descriptions, and $R$ indicates a case (role)
description.
Descriptions are determined using the syntax as
in Figure~\ref{pic:0}.
\begin{figure*}
\[
\begin{array}{rrl}
{\rm primitive\ concept}       & C_1, C_2 \to & A | \\
{\rm conjunction,\ disjunction} && C_1 \sqcap C_2 | C_1 \sqcup C_2 | \\
{\rm negation}&& \neg C | \\
{\rm universal\ quantifier} && \forall R.C | \\
{\rm existential\ quantifier} && \exists R.C | \\
{\rm number\ restrictions} && (< n R) | (\le n R) | (= n R) |
                              (\ge n R) | (> n R)  \\
{\rm case\ conjunction} & R \to & P_1 \sqcap \dots \sqcap P_m
\end{array}
\]
\caption{Syntax of descriptions. } \label{pic:0}
\end{figure*}

\subsubsection{Semantics}

Semantics of the constructions is given by the {\em assignments}
$I$ that is, in the context of this subsection, the same as {\em
interpretations}. We assume that $I$ is being associated a
non-empty {\em domain} $H(I)$. Note, that best of all think of the
term `domain' in a sense of the theory of
computations~\cite{Sco:71}, \cite{Sco:80}. This interpretation
assigns a unary relation $A(I)$ over $H(I)$ to each atomic
(primitive) concept $A$ as well as a binary relation $R(I)$ over
$H(I) \times H(I)$ to every atomic (primitive) case $R$.

In the following $card\{S\}$ means the cardinality
of a set $S$, binary relation $\theta$ is one of $\{<, \le, =, \ge, > \}$,
$C_i \subseteq H(I)$,
 and the informal ideas are determined
by the equations shown in Figure~\ref{pic:1}.
\begin{figure*}
\[
\begin{array}{rcl}
(C_1 \sqcap C_2)(I) & = & C_1(I) \cap C_2(I),\\
(C_1 \sqcup C_2)(I) & = & C_1(I) \cup C_2(I), \\
(\forall R.C)(I) & = & \{h \in H(I)|
                          \forall d : (h,d) \in R(I) \Rightarrow
                          d \in C(I) \}, \\
(\exists R.C)(I) & = & \{h \in H(I)|
                          \exists d : (h,d) \in R(I) \land
                          d \in C(I) \}, \\
(\theta\ n\ R)(I)  & = & \{h \in H(I)|
                          card \{d : (h,d) \in R(I)\}\ \theta\ n \}, \\
(P_1 \sqcap \dots \sqcap P_m)(I) & = &
                P_1(I) \cap \dots \cap P_m(I).
\end{array}
\]
\caption{Semantics of descriptions. } \label{pic:1}
\end{figure*}

\section{Event driven objects}\label{event-driven-objects}

A description logic system provides some services
that are connected to the computational features
of the event-driven environment.

First of all it provides the procedures for validating
the subsumptions between concepts. For any two concept
descriptions it can validate whether one of the descriptions
{\em for all assignments} $I$, i.e. {\em always}, determines
a superset of the {\em individuals} for the other.
Say, the derived concept $C_1$ is subsumed by the
concept {\tt author} as follows:
\begin{center}
{\tt author := person $\sqcap$ (> 0 paper)}, and \\
$\forall I$.{\tt author(I) := person(I) $\sqcap$ (> 0 paper)(I)}
\end{center}

The term `event driven object' means that the {\em script} is executed
in response to the {\em event} being recognized by the object.
Every event has the associated script.

To enable the event-driven computations,
the Modular Counterpart (MC) is implemented as
a holder of all the controls to communicate with the user.

\begin{figure*}
\centerline{
\unitlength=1.00mm \special{em:linewidth 0.4pt}
\linethickness{0.4pt}
\begin{picture}(52.00,60.00)
\put(23.00,53.00){\makebox(0,0)[cc]{{\bf P}otential {\bf
O}bjects}} \put(21.00,15.00){\framebox(10.00,32.00)[cc]{}}
\put(26.00,43.00){\circle{2.00}} \put(26.00,33.00){\circle{2.00}}
\put(26.00,23.00){\circle{2.00}}
\put(23.00,36.00){\makebox(0,0)[cc]{h}}
\put(3.00,39.00){\makebox(0,0)[cc]{{\bf E}vents}}
\put(1.00,1.00){\framebox(10.00,32.00)[cc]{}}
\put(6.00,29.00){\circle{2.00}} \put(6.00,19.00){\circle{2.00}}
\put(6.00,9.00){\circle{2.00}}
\put(3.00,22.00){\makebox(0,0)[cc]{i}}
\put(7.00,20.00){\vector(2,1){40.00}}
\put(44.00,60.00){\makebox(0,0)[cc]{{\bf A}ctual {\bf O}bjects}}
\put(42.00,22.00){\framebox(10.00,32.00)[cc]{}}
\put(47.00,50.00){\circle{2.00}} \put(47.00,40.00){\circle{2.00}}
\put(47.00,30.00){\circle{2.00}}
\put(46.00,43.00){\makebox(0,0)[cc]{h(i)}}
\end{picture}
}
\caption{ {\bf E}vent {\bf D}riven {\bf O}bjects.
{
\em Here: possible object $h$ is the mapping
from the event (assignment) $i$ into the actual object $h(i)$.
Note that a set of all the possible objects $\{h|h:I \to T\}=H_T(I)$
represents an idea of functor-as-object for $I$ is a category
of events, T is a (sub)category of the actual objects - type.
}
}\label{pic:A}
\end{figure*}
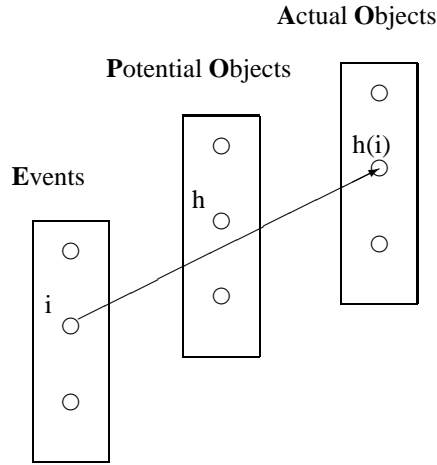

The event -- and corresponding {\em assignment}, or
interpretation, -- is assigned by the user call (for instance,
clicking) or selection. Thus, when the activity is initiated, the
following main events may be triggered: respond to a request from
the user application, database retrieval or updating. The possible
order of the events is prescribed by {\em evolver} and is
determined by the {\em scripts}. A fragment of the event driven
procedure is shown in Figure~\ref{pic:A}.

Semantic heterogeneity is the result of representing the same
or overlapping data in two or more ways.
The ways to compare the ability of different data models
and database schemas to hold the same information, possibly
restructured, could be derived.

As usually, events could be triggered using {\em menu} in the user GUI,
initiating the associated {\em particular application}.

Menu gives more flexibility to the attribute selection. Usually
the lists of possible attributes are supported to give the
developer or user more freedom. Menus are established to be
encapsulated in {\em Ancestor Potential Objects} (APO) and are
inherited from {\em Descendant Potential Objects} (DPO).

The particular application is derived from Potential Object Library (POL)
giving rise to Actual Object Libraries (AOL).

\subsection{The starting assumptions}

The represented domain is assumed inhabited
by the (atomic) entities, or {\em individuals}. A safety reason
is to set up individual as a primary concept that is not assumed
to be definable. In fact, the observer operates with the
{\em constructs} that represent the individuals that
can be located into a single domain $D$.

The advanced studies in a theory of computations prescribe $D$
as a domain of {\em potential} (or {\em schematic}) individuals.
Those individuals are possible
with respect to some theory (of individuals).

The individuals enter the domain and leave it
cancelling out their own existence. Such a `flow of events'
may be based on a time flow.

The additional {\em virtual} individuals
are completely {\em ideal objects}.
They are used to increase the structure
regularity of the (initial) domain $D$.

As a result, clear distinction between {\em actual, possible}
and {\em virtual} individuals induces the inclusion:

\[ A \subseteq D \subseteq V, \]

where $A$ is a set of actual individuals,
$D$ is a set of possible individuals,
and $V$ is a set of virtual individuals.
The central computational proposal is to
generate actual individuals as the different {\em families} of
$D$,

\[ A_i \subseteq D\ {\rm for\ } i \in I.  \]

\subsection{Other generic notions}

A user actually needs a (logical) language, even
overcoming his own initial desire.
These logics is not homogeneous and
 do not suit the
amorphous idea of a thing and property. The regular and working
logics are the logics of {\em the descriptions}. The descriptions
directly illustrate the differences of the individuals and tend to
general operators.

The logical formula $\Phi (x)$ gives
the property, but the direct assignment of the property
$\Phi (\cdot)$ to the individual $x$ is given by
the description:

\[ {\cal I}x.\Phi (x),  \]

with a sense `the (unique) $x$ that $\Phi (x)$'.

The connection between syntax and semantics
  is given by the {\em evaluation map}:
$$
\parallel \cdot \parallel \cdot\ : \
{\rm descriptions\ }\times\ {\rm assignments\ }
\rightarrow\ {\rm individuals}.
$$
(Here: an assignment is temporary viewed as an index
ranging the families.)
The abridged concepts are an {\em attribute} $a$ and
{\em property} $\Phi (\cdot)$ (via the description):

$$
a = \parallel {\cal I}x.\Phi(x)\parallel_i\ {\rm for\ } i\in I
\eqno (\bf Attr)
$$

An attribute thus defined indicates the set of individuals
with a property $\Phi(\cdot)$. In usual terms the
{\em functional representation of attribute} is established
(attribute is a mapping {\em from} a set of things
and a set of `observation points' {\em into} a set of values).
Note that a `thing' is represented by the `description'.

\begin{namelist}{{\tt -h} {\it {   }}}
\item {\sf Principle adopted}: The attribute is defined
by ({\bf Attr}). The addition of the uniqueness

$$
\{ a \} = \{ d \in D \mid\ \| \Phi(\bar{d})\|_i
= true \} \eqno  ({\bf Singleton})
$$

as necessary and sufficient condition
$$
\| {\cal I}x.\Phi(x)\|_i = a
  \Leftrightarrow \{ a \} =  \{ d \in D \mid\ \| \Phi(\bar{d})\|_i
        = true \} \eqno  ({\bf Unique})
$$

enforces the observer to conclude: fixing the family $i\in I$
and evaluating $\|\Phi(\bar{d})\|_i$ relatively to every $d\in
D$,
he verifies the uniqueness of $d$.
\end{namelist}

In above the individual is called as $a$ and is adopted as
an evaluation of the description relatively to $i$.

\subsection{Functional scheme}

A general solution for attributes attracts the set of attribute
functions ({\bf Attr}) that is called as a {\em functional
scheme}.

Equation ({\bf Attr}) is to be revised
as follows:
\[
\begin{array}{ll}
{\cal I}x.\Phi(x) = \bar{h} & {\rm in\ a\ language\ of\ observer}
\vspace{0.5ex} \\

\|\bar{h}\|=h              & {\rm this\ is\ an\ individual} \\
                           & {\rm concept\ in\ a\ domain}
\vspace{0.5ex}                                               \\

h(i)=a                     & {\rm this\ is\ an\ individual} \\
                           & {\rm in\ a\ domain}
\end{array}
\]

Thus, if $h$ is an individual, then $a$ is its {\em state}
under the {\em forcing condition} $i$.

Hence, the generalized individuals (or: concepts)
are schematic:

\[ h : I \rightarrow C, \]

where $h$ is a mapping from the `observation points' into
the (subset of) attribute $C$. The latter undoubtedly
is {\em the} set of individuals.

There is a clear reason to call the collection of $h$
as a {\em concept}.
Thus a concept really represent the functional scheme.
The (individual) functional schemes
are to be gathered into a greater stock:
$$
\{h \mid h: I \rightarrow C\} = H_C(I)  \eqno  ({\bf VDom}).
$$

Certainly, $H_C(I)$ is and idealized object.
This object $H_C(I)$ is
a representation, and what is specific
the feature of a {\em variable domain} is captured.
The possibilities and the advantages of a notion of
variable domain are applied mostly to the {\em dynamics}.

\subsection{Dynamics of objects}

The state in an object-oriented approach is viewed as the value
of the functions in the functional scheme at a given point
among the `observation points'. This agrees with the
computational framework where.
a set of individuals
is generated by:

\[ H_C(\{ i\} ) \subseteq C\ {\rm for\ } i\in I.  \]

This set is a state of a variable domain $H_C(I)$, where
$C$ gives the local universe of possible individuals.
The pointer $i$ marks the family of individuals that
is `observed' from $i$.

As can be shown in addition, the commonly used in object studies are
{\em encapsulation, composition, classification}, and
{\em communication/transaction} have the computational
representation as well.

\section{Integrating data}\label{integrating-data}

The desired aim for data integration could be applied to
constructing a global schema from the source database schemas. In
the project under discussion there are two ways to supported {\em
read-only} integrated access via views: {\em virtual} and {\em
actual}.

\begin{description}
\item[The virtual approach.] This way is based on
      decomposing the initial query to subqueries
      being addressed to particular source databases.
\item[The actual approach.] The main feature is that
      a view is updated every time when given an update
      in terms of particular databases. This is important
      for commercial applications when data warehouse provide
      support to actual integrated view.
\end{description}

ADE has under research the idea of a concept as the
variable entity to
possess the creation of the variable concepts
and associated transition effects.
In their turn the variable concepts lead to
parameterized type system. The
approach developed in ADE is based on the reasons stated.

     The usage of the method of embedding typed system (including the
apparatus of variable concepts) into untyped system based on the apparatus
of Applicative Computational Systems (ACS) is the distinctive feature of the
approach being developed.

     Combining the ideas of variable concepts will make possible
development of a wide range of applied information systems, particularly in
the field of data base management systems, knowledge based systems and
programming systems.

ADE is viewed to be a comprehensive computational
environment from a formal point of view based on the notion
of  `variable concept'.
This notion gives rise to an approach to integrating
the far distant concepts, means and
models in use.

The target prototype system Application Development Environment (ADE) is
involves the idea of a variable, or switching concept and covers the
vital mechanisms of encapsulation, inheritance and polymorphism.
Variable
concepts naturally generate families of similar types
that are derived from the
generic types.
Concepts in ADE are equipped with the evolvents that manage
the transitions, or switching between the types.

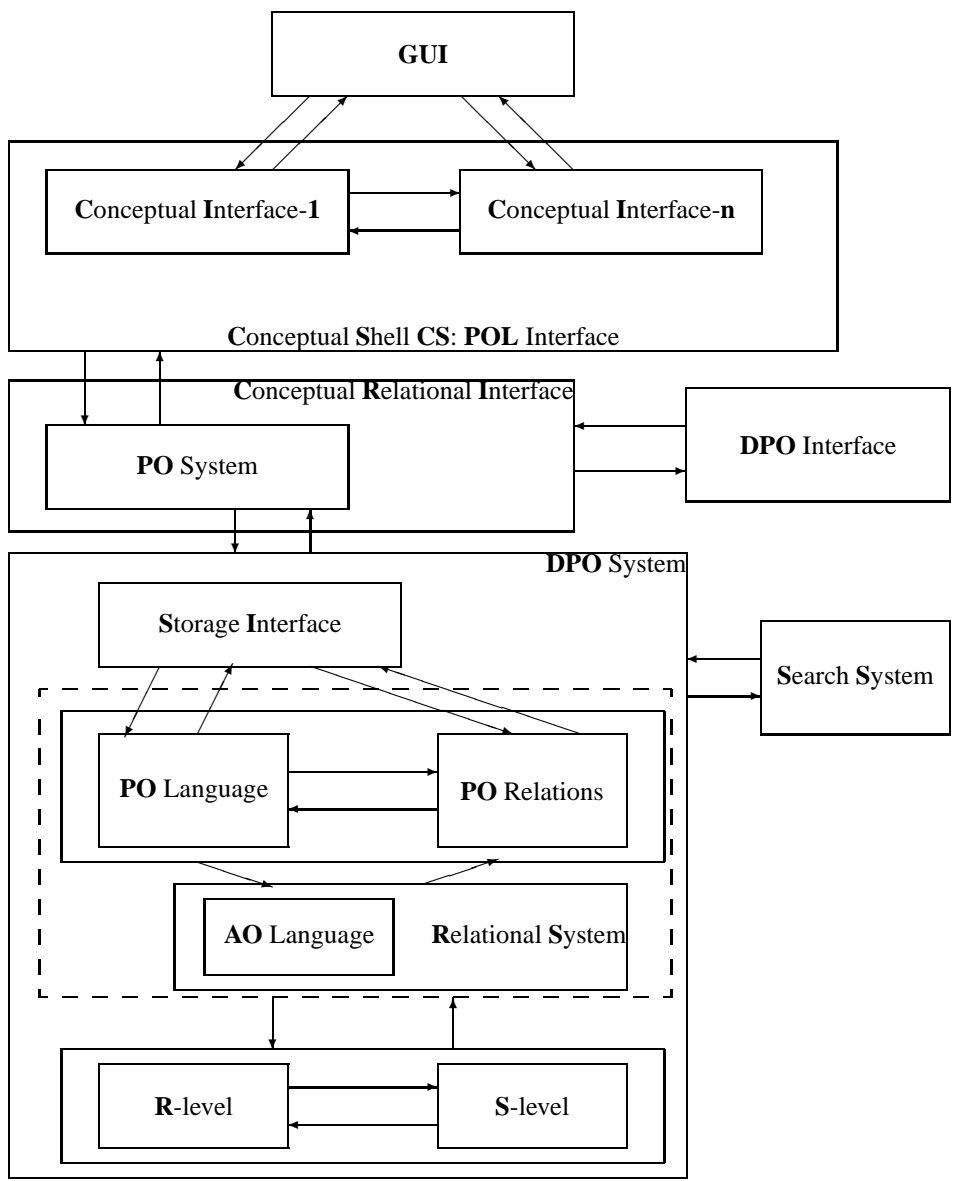
\begin{figure*}
\begin{center}
\unitlength 1.00 mm \linethickness{0.4pt}
\begin{picture}(126.00,156.00)
\put(36.00,145.00){\framebox(40.00,11.00)[cc]{{\bf GUI}}}
\put(6.00,90.00){\framebox(40.00,11.00)[cc]{{\bf PO} System}}
\put(91.00,91.00){\framebox(35.00,15.00)[cc]{{\bf DPO} Interface}}
\put(13.00,69.00){\framebox(40.00,11.00)[cc]{{\bf S}torage {\bf
I}nterface}} \put(13.00,45.00){\framebox(25.00,15.00)[cc]{{\bf PO}
Language}} \put(58.00,45.00){\framebox(25.00,15.00)[cc]{{\bf PO}
Relations}} \put(27.00,28.00){\framebox(25.00,10.00)[cc]{{\bf AO}
Language}} \put(13.00,5.00){\framebox(25.00,11.00)[cc]{{\bf
R}-level}} \put(58.00,5.00){\framebox(25.00,11.00)[cc]{{\bf
S}-level}} \put(8.00,3.00){\framebox(80.00,15.00)[cc]{}}
\put(1.00,111.00){\framebox(110.00,27.00)[cb] {{\bf C}onceptual
{\bf S}hell {\bf CS}: {\bf POL} Interface}}
\put(8.00,43.00){\framebox(80.00,20.00)[cc]{}}
\put(1.00,87.00){\framebox(75.00,20.00)[rt]{{\bf C}onceptual {\bf
R}elational {\bf I}nterface}}
\put(23.00,26.00){\framebox(60.00,14.00)[rc]{{\bf R}elational {\bf
S}ystem}} \put(1.00,1.00){\framebox(90.00,83.00)[rt]{{\bf DPO}
System}} \put(6.00,124.00){\framebox(40.00,11.00)[cc]{{\bf
C}onceptual {\bf I}nterface-{\bf 1}}}
\put(61.00,124.00){\framebox(40.00,11.00)[cc]{{\bf C}onceptual
{\bf I}nterface-{\bf n}}} \put(36.00,135.00){\vector(1,1){10.00}}
\put(61.00,145.00){\vector(1,-1){10.00}}
\put(46.00,132.00){\vector(1,0){15.00}}
\put(61.00,127.00){\vector(-1,0){15.00}}
\put(91.00,101.00){\vector(-1,0){15.00}}
\put(76.00,95.00){\vector(1,0){15.00}}
\put(31.00,90.00){\vector(0,-1){6.00}}
\put(41.00,84.00){\vector(0,1){6.00}}
\put(76.00,135.00){\vector(-1,1){10.00}}
\put(41.00,145.00){\vector(-1,-1){10.00}}
\put(21.00,69.00){\vector(-1,-2){4.67}}
\put(26.00,60.00){\vector(1,2){4.67}}
\put(38.00,55.00){\vector(1,0){20.00}}
\put(58.00,50.00){\vector(-1,0){20.00}}
\put(38.00,13.00){\vector(1,0){20.00}}
\put(58.00,8.00){\vector(-1,0){20.00}}
\put(11.00,111.00){\vector(0,-1){10.00}}
\put(21.00,101.00){\vector(0,1){10.00}}
\put(101.00,60.00){\framebox(25.00,15.00)[cc]{{\bf S}earch {\bf
S}ystem}} \put(101.00,70.00){\vector(-1,0){10.00}}
\put(91.00,65.00){\vector(1,0){10.00}}
\put(5.00,25.00){\dashbox{2.00}(84.00,41.00)[cc]{}}
\put(26.00,43.00){\vector(3,-1){10.00}}
\put(56.00,40.00){\vector(3,1){10.00}}
\put(36.00,25.00){\vector(0,-1){7.00}}
\put(60.00,18.00){\vector(0,1){7.00}}
\put(41.00,69.00){\vector(3,-1){27.00}}
\put(77.00,60.00){\vector(-3,1){27.00}}
\end{picture}
\end{center}
\caption{ {\bf A}pplication {\bf D}evelopment {\bf E}nvironment: {\bf ADE}.
{
\em     Abbreviations:
         {\bf GUI} - {\bf G}raphical {\bf U}ser {\bf I}nterface;
         {\bf POL} - {\bf P}otential {\bf O}bject {\bf L}ibrary;
         {\bf PO}  - {\bf P}otential {\bf O}bject;
         {\bf AOL} - {\bf A}ctual {\bf O}bject {\bf L}ibrary;
         {\bf AO}  - {\bf A}ctual {\bf O}bject;
         {\bf DPO} - {\bf D}escendent {\bf P}otential {\bf O}bject;
         {\bf R}-level - {\bf R}epresentation level;
         {\bf S}-level - {\bf S}torage level
}
}\label{pic:B}
\end{figure*}

In particular, the identity evolvent supports the constant
concepts and types (statical concepts). To achieve the needed
flexibility a general ADE layout consists of the uniform modular
units, as shown in Figure~\ref{pic:B}.

In ADE Data Object Definition Language (DODL) contains the construction
of data objects' base schema as a relation between concepts.
Concepts are
included into the type system with the interpretation over the variable
domains. A coherent set of variable domains generates the data objects'
base. Basis to maintain the data objects in use and their bases
is generated by
computational models with applicative structures.

The developer obtains the set of the means that establish, support and modify
the linkages between the data objects' base schemes, data objects' base and
computational models. DODL declares: {\em type system} as a set of metadata
objects; em linkages between the types; system of domains;
linkages between the
domains; extentions of domains and types; computational tools of applicative
pre-structures and structures.

The third part of the implementation supports two level of
interfaces. The first is the Intentional Management System (IMS)
to support concepts (metadata objects) of different kinds, and the
second is associated Extensional Management System (EMS) to
support the appropriated extensions (data objects) generated by
the intentions.

EMS is embedded into the unified computational model. It is
object-oriented extensible programming system Basic Relational
Tool System (BRTS). BRTS has the fixed architecture with the one
level comprehension, separate self- contained components,
interfaces and languages. It is the First Order Tool (FOT) and
generates `fast prototypes'. D(M)ODL and D(M)OML of BRTS contain
the SQL-based relational complete languages that cooperate with
ADE. BRTS mainly supports relatively large number of low
cardinality relations (extensions) and supports Data (Metadata)
Object Model D(M)OM with retrieval, modifications and definitions
of a storable information.

IMS is also embedded into the computational model and supports a numerous
matadata objects. Their amount is almost the same as for data objects. IMS
is based on D(M)OM with a simple comprehension to manage metadata base
and is supported by MetaRelational Tool System (MRTS). MRTS
manipulates with the metaobjects (concepts) and metarelations (frames) and
is embedded into ADE.

\section{Supporting technologies}\label{supporting-technologies}

A main result is the experimental verification
of variable concepts approach.
This would be applied to develop the variety of
applied information systems.

Computations with variable concepts and appropriated programming
system allows to built a system especially to
manipulate the objects.

The experimental research and verification of the obtained
model is based on prototypes -- CS, ADE, BRTS, BMRS.
The difficulties to
implement full scale prototype are resolved by the high level object-oriented
programming language. Some candidate programming systems are tested to
enable the needed computational properties. After that the main
programming tool kit is selected. Preliminary candidate tools were C++ or
Modula-2. An attention is paid to select an appropriate database management
system. If needed the original DBMS is attached.
At the preliminary tests the attention was paid to OLE-2 techniques.

Some ready made original systems were tested and expanded to
achieve the prototype system with the properties
mentioned.

\section*{Conclusions: interpretation of the results}
\addcontentsline{toc}{section}{Conclusions}

Note that our goal is to obtain close to the same efficiency as
would have been offered by a custom-built DL reasoner.
Of course, the approach presented here is not a perfect.
\begin{itemize}
\item[1)] To the extent that normalize­compare
algorithms are unable to
reason in a complete manner with DL constructors
involving incomplete knowledge such as
disjunction, the present system is also likely to suffer
the same deficiencies.
\item[2)] The present work has not yet addressed DL notions
such as role constructors, recursive concepts,
and general constraints.
\item[3)] There are many other notions in
knowledge representation, such as the full spectrum of
epistemic and other non­monotonic
reasoning, abduction, case­based reasoning, etc.,
which are likely to require a thorough
overhaul of the entire reasoning architecture,
and hence are likely not to be accommodated
properly by the present approach.
\end{itemize}
The resulting two level comprehension model and
computational environment verify the feasibility of the approach.
The adequate, neutral and semantical
representation of data is the target in the sphere of
extensible systems and
their moderations and modifications.
The relational solutions are the criteria in database technology.
Therefore, the variable concepts generate the
power and sound representation of data
objects, have the boundary conditions as the known
results in information
systems (both in a theory and applications) and capture
the additional effects
of dynamics to simulate,
in particular, the encapsulation, polymorphism and
inheritance.

\nocite{Rou:76}
\nocite{Brod:95}
\nocite{OMG1:91}
\nocite{KaWi:96} \nocite{HeSa:95} \nocite{And:96}
\nocite{Wo:96}
\nocite{BeLeRo:97} \nocite{Hull:97}
\nocite{Bor:95} \nocite{Bor:96}
\nocite{Wo:99}

\addcontentsline{toc}{section}{References}
\newcommand{\noopsort}[1]{} \newcommand{\printfirst}[2]{#1}
  \newcommand{\singleletter}[1]{#1} \newcommand{\switchargs}[2]{#2#1}


\end{document}